\documentclass[12pt]{iopart}
\usepackage{graphicx}
\usepackage{citesort}  
\usepackage{amssymb}
\usepackage{tabulary}
\usepackage{tabularx}
\usepackage{multirow}
\usepackage{color}

\begin{document}
\bibliographystyle{agsm} 
\title[First-Principles High Pressure Studies on Group-14 Element Pernitrides]{First-Principles High Pressure Studies on Group-14 Element Pernitrides}

\author{Sharad Babu Pillai$^1$, Himadri R. Soni$^{2*}$ and Prafulla K. Jha$^1$}

\address{$^1$ Department of Physics, Faculty of Science, The M. S. University of Baroda, Vadodara-390002, India}
\address{$^2$ School of Sciences, Indrashil University, Rajpur, Kadi 382740, India}
\ead{*himadri.iu@gmail.com}


\begin{abstract}
The search of ultra hard materials is inevitable in high pressure device applications. Nitrides of group-14 elements have been foreseen as potential candidates in replacing existing hard materials. In a recent experiment, pyrite structures of SiN$_2$, GeN$_2$ and SnN$_2$ have been synthesized at high pressure and were shown to have high bulk modulus. Though their existence and bulk modulus are known, limited studies have been devoted to SiN$_2$, GeN$_2$ and SnN$_2$. In the present work, we have performed the first-principles calculations to investigate structural, electronic, mechanical and vibrational properties at ambient as well as high pressure condition. The physical properties of SnN$_2$ and high pressure lattice dynamical properties of SiN$_2$ and GeN$_{2}$ are explored for the first time. SiN$_2$ has higher bulk modulus among all  MN$_2$  (where M = Si, Sn and Ge), and increases further with increase in pressure. The increase in elastic moduli of MN$_2$ have been related to the shortening of M-N bond length at high pressure. Electronic properties of these pyrite structures suggest that the bandgap increases with applying pressure. We further characterize these MN$_2$ at high pressures using theoretically calculated Raman spectroscopy. Frequency of N-N stretching A$_g$ and T$_g$ modes confirms the single bond character of nitrogen dimer in all MN$_2$.
\end{abstract}

%
\vspace{2pc}

\section{Introduction}

The search for excellent mechanical properties such as high hardness and strength of a material has drawn the attention of researchers towards ultrahard and superhard materials like diamond and cubic boron nitride \cite{132014haines2001}. However, due to their own limitations, new materials such as nitrides, oxides, and borides with properties comparable to those of diamonds \cite{142014firstSproul889,152014firstKUMAR2012150,162014firstMOSUANG2002363,172014firstLowther_2005} are being searched for better replacements.
Exploration of pernitrides began after many experimental and theoretical attempts in search of nitrogen rich stoichiometries  \cite{92017theo,102017theo,112017theo,122017theo,132017theo,142017theo,152017theo,162017theo,172017theo,182017theo,2017theo20171179}. A series of transition metal pernitrides (Mn$_2$, Ta$_2$, Nb$_2$, V$_2$, Zr$_2$, and Hf$_2$,) were predicted using combined first-principles calculations and global structure search method. Recently, Alkhaldi and
Kroll \cite{2019alkhadi} considered chemical potential change of molecular nitrogen in their first-principles calculations and they further derived pressure-temperature phase diagrams for guiding experimental synthesis as well as predicting the  formation of nitrogen rich stoichiometries under high pressure and temperature. The quest for nitrogen rich materials is still considered as an important motif among material scientists due to their interesting chemical properties and possibility of being environmental friendly high density material following which researches have proposed various experimental and conceptual strategies for having nitrogen rich compounds, \cite{72018Vajenine2001,82018,92018,102018Bykov2018,112018PhysRevMaterials.2.011602,Li,132018doi:10.1021/acs.chemmater.6b00042,142018doi:10.1021/acs.jpcc.7b02592,152018doi:10.1021/acs.jpcc.5b08595,162018STEELE201621,172018doi:10.1002/anie.201702440,182018doi:10.1021/acs.chemmater.6b00042,192018-expC7DT01583F,202018doi:10.1021/acs.inorgchem.7b00516,2018acsbarium} including stable nitrides with electro-positive elements at high pressure \cite{Li,2018acsbarium}. Several group-14 nitrides of the stoichiometry M$_3$N$_4$, where M replaces the group-14 elements, have been widely investigated for their mechanical properties both experimentally and theoretically \cite{18014firstliu1989,192014firstlowther2000,202014firstching2001,212014firstzerr1999,222014firstshem2002,232014firstserghiou1999}. The high stability and easy release of molecular N$_2$ makes the synthesis of nitrogen-rich stoichiometries such as C$_3$N$_4$ and C$_1$$_1$N$_4$ challenging\cite{2002WEIHRICH2003701}. Weihrich \etal \cite{2002WEIHRICH2003701} provided a hypothetical pyrite type structure for CN$_2$, iso-electronic to SiP$_2$ \cite{chattopadhyay1984} and proposed that the formation of CN$_2$ does not involve the breakage of strong N-N bond of N$_2$ compared to the formation of other CN$_x$ compounds. The realization of the pyrite structure allows building a three-dimensional network of C-N bonds without changing the strong N-N bonding.

Limited studies have been done on group 14 elemental pernitride stoichiometry MN$_2$ (M=Si, Ge, Sn) in pyrite structure \cite{2003weihrich2003,2011DING20111357,2014rscC4RA11327F}. Weihrich \etal \cite{2003weihrich2003} showed the pyrite SiN$_2$ to be an insulator with calculated optical band gap of 5.5 eV and GeN$_2$ to be semiconducting with band gap of 1.4 eV. Chen \etal \cite{2014rscC4RA11327F} suggested the formation of SiN$_2$ in the pyrite structure at a pressure above 17 GPa and showed its super hard nature and predicted the lattice dynamical stability upto 100 GPa. They also calculated the elastic properties of pyrite SiN$_2$ and proposed that the bulk modulus can reach upto 320 GPa indicating ultra incompressibility. Manyali \etal \cite{2014elsevierMANYALI2014706} studied the structural, electronic and optical properties of the SiN$_2$ and GeN$_2$. However, they considered them in orthorhombic structure adopted from Copper Silver Sulphide (CuAgS). SiN$_2$ and GeN$_2$ in orthorhombic structure are indirect band gap semiconductors. It was only recently that the pyrite form of pernitrides of Si, Ge, and Sn were synthesized at high pressure through a direct chemical reaction between group-14 elements and molecular nitrogen \cite{2017experimental}. The experimentally determined bulk modulus of SiN$_2$, GeN$_2$ and SnN$_2$ is 354, 284 and 219 GPa, respectively, which is higher than those of the corresponding spinel phases \cite{122017pyritePhysRevB.65.161202,132017pyriteshemkunas_petuskey_chizmeshya_leinenweber_wolf_2004,152017experimentalPhysRevB.82.144112}.The mechanical and vibrational properties of GeN$_2$ and SnN$_2$ remain unexplored and those for SiN$_2$ are limited. Therefore, a systematic investigation of the physical properties of these nitrogen rich stoichiometries (SiN$_2$, GeN$_2$ and SnN$_2$) is essential. Furthermore, being nitrides of group-14 elements, pyrite structure of SiN$_2$, GeN$_2$ and SnN$_2$ deserves extra attention for the study of mechanical properties. The hardness of a material can be increased by increasing the density and altering the electronic bonding states of materials \cite{72014-thestabilityZhao2014}, for which pressure can be an effective tool. Hence, the study of mechanical properties of superhard materials at high pressure is inevitable. Among many judging parameters, bulk and shear moduli are important properties to consider for potential ultrahard materials.

Therefore, in this work, we have performed  detailed investigation of structural, electronic, mechanical and vibrational properties of nitrogen rich MN$_2$ (where M = Si, Ge and Sn) stoichiometries in their pyrite structure at ambient and high pressure conditions. Among the studied compounds, no theoretical studies have been done on the pyrite structure of SnN$_2$ to the best of our knowledge. Further, we relate the high pressure stability of these pernitrides with the electronic properties and chemical bonding. Our study provides new insights into physical properties of group-14 pernitrides, which have not been studied for these compounds at high pressure so far.

\section{Computational Details}

The ambient and high pressure structural, electronic, mechanical and vibrational properties of MN$_2$ were studied using density-functional theory (DFT) based on first-principles calculations \cite{dftPhysRev.140.A1133}. Structural optimization and total energy calculations were carried out using Quantum Espresso \cite{quantumespresso}, which implements DFT to solve the Kohn-Sham equations. We used both local density approximations (LDA) \cite{ldapwPhysRevB.45.13244} and generalized gradient approximations (GGA) \cite{ggapbePhysRevLett.77.3865} to treat the exchange-correlation functional for determining the structural, electronic and mechanical properties of MN$_2$ at ambient condition. Analyzing the degree of closeness of the calculated structural parameters with experimental values, and previous theoretical studies showed that the structural and mechanical properties of pyrite group-14 pernitrides are better estimated by LDA, whereas, GGA provides better estimates of electronic band gaps. Hence, for high pressure studies, we used LDA for mechanical properties and GGA for electronic band structure and phonon dispersion calculations. The core electrons of all atoms were treated using norm-conserving pseudopotential under Martins-Troullier method \cite{martinstrouierPhysRevB.43.1993}. To expand the one electron orbital and electronic charge density, plane wave basis sets were used with kinetic energy cutoff set to 150 Ry for all considered systems. Brillouin zone integration was performed over  $6 \times 6 \times 6$  \textbf{k}-point grid generated under Monkhorst-pack scheme \cite{MONKHORSTPhysRevB.13.5188}. To achieve the total energy convergence for the considered geometries, the forces on atoms were set to a threshold value of 0.0001 Ry.  To calculate the elastic constants and elastic moduli, the geometry of considered systems at each pressure was deformed with the maximum absolute value of Lagrangian strain set to 0.15 producing 11 distorted structures with the deformation type implemented in Elastic - The exciting code \cite{elasticGOLESORKHTABAR20131861}. For each deformed structure, internal degrees of freedom were optimized and the total energy was evaluated using Quantum Espresso package \cite{quantumespresso}. The energy-strain curve thus obtained was polynomial fitted to obtain the second derivative of energy with respect to Lagrangian strain at equilibrium. The coefficient of the quadratic term of a best-fitted polynomial can be expressed as a linear combination second-order elastic constants. With such procedure, each deformed structure yields a set of linear equations, which were solved using linear square fit to get second order elastic constants. Elastic moduli were calculated using the Voigt and Reuss approach \cite{Chung1967}. Since Voigt and Reuss elastic moduli were shown to be strict upper and lower bound respectively, we calculated the Hill averaged values of elastic moduli of considered systems. The lattice dynamical properties were calculated using density-functional perturbation theory (DFPT)\cite{dfptRevModPhys.73.515}. Phonon frequencies were obtained by Fourier transform of dynamical matrix on $6 \times6 \times 6$ grid of \textbf{q-} points. Raman activity of phonon modes were determined implementing the second order response method \cite{ramanPhysRevLett.90.036401}.

\section{Results and Discussion}

\subsection{Structural Properties}

\begin{figure}
\centering
\includegraphics[width=0.5\textwidth]{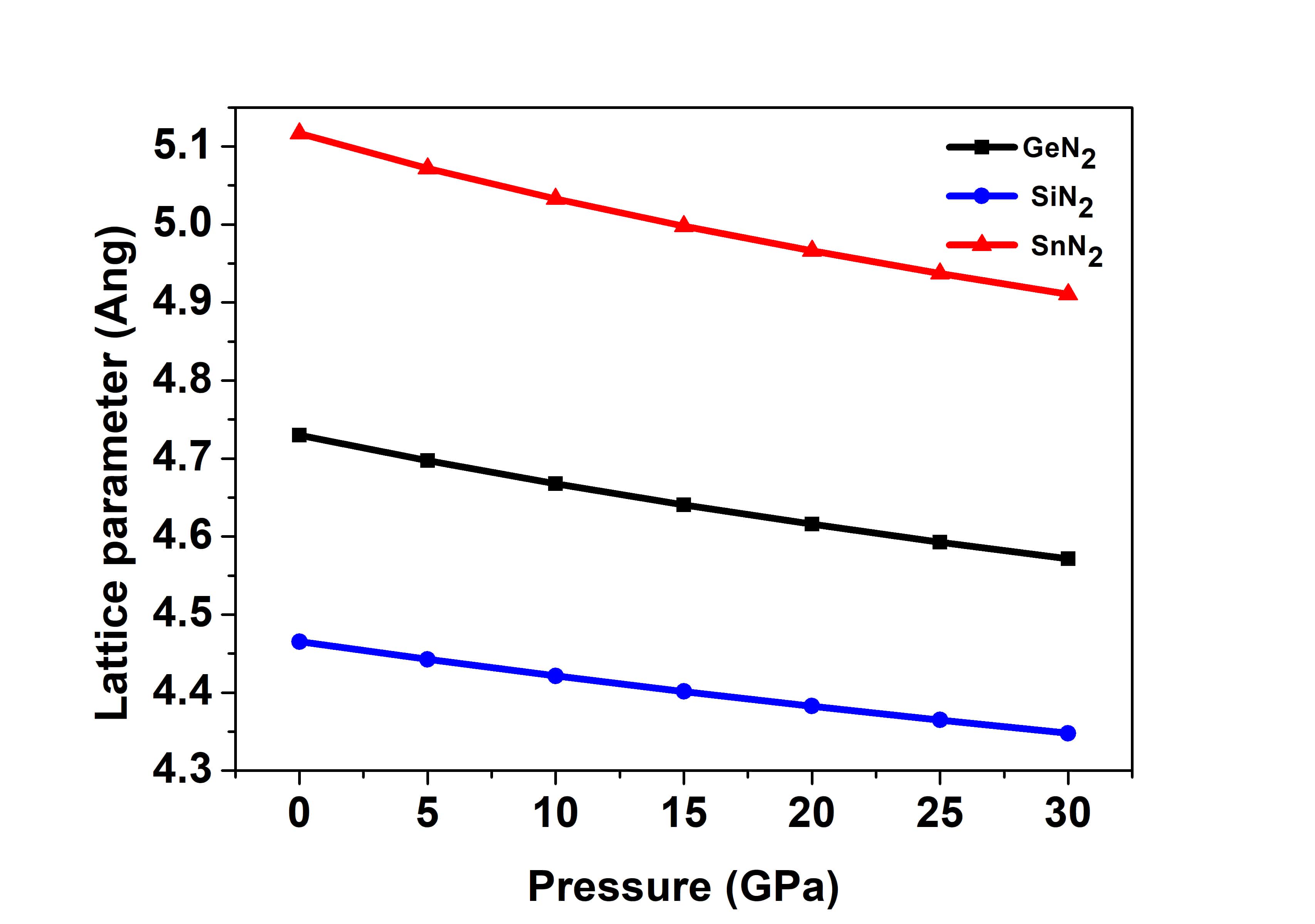}
\includegraphics[width=0.5\textwidth]{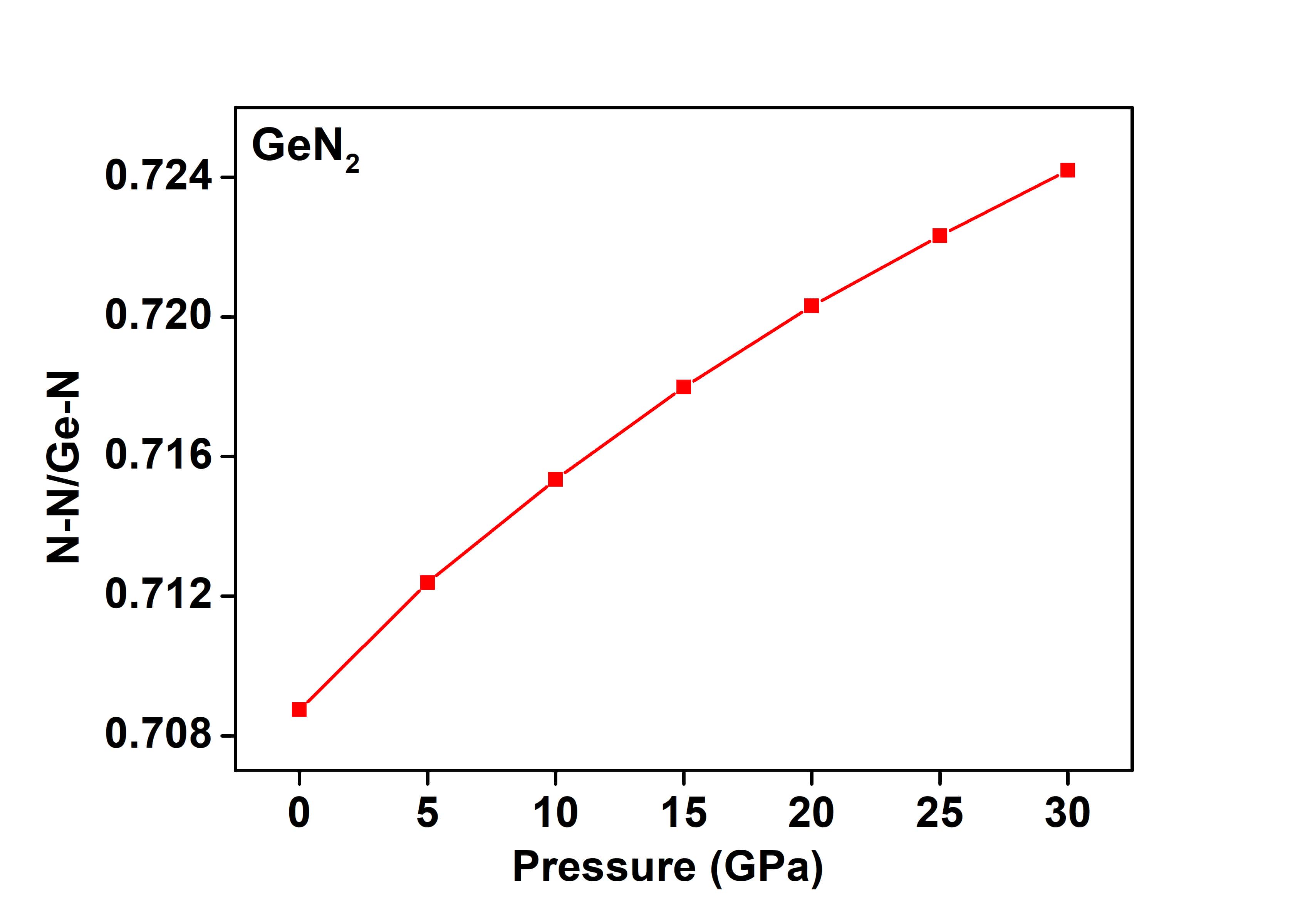}
\caption{Variation of lattice parameter (top) and d$_{N-N}$/ d$_{M-N}$ (bottom) versus pressure.}
\label{fig:lattice}
\end{figure}

Pyrite structures with nitrogen stoichiometry MN$_2$ (M = Si, Ge, Sn) have 'M' atom residing at the centre of octahedral formed by six nitrogen atoms. A recent experimental study reports the crystal structure of these pernitrides synthesized at 60 GPa pressure through a direct chemical reaction between group-14 elements and N$_2$ \cite{2017experimental}. Prior to high pressure study, we re-investigated the crystal structure of MN$_2$ in pyrite structure to support the experimental findings as well as for exploring their high pressure behaviour. We initially performed structural relaxation of MN$_2$ in pyrite structure, and the optimized lattice parameter, bond lengths and bond angles are presented in Table.~\ref{tab:example_1} along with available experimental\cite{2017experimental} and theoretical data \cite{2003weihrich2003,2011DING20111357,2014rscC4RA11327F}. Both LDA and GGA functionals were tested for calculations and were compared with the available results wherever applicable. It is evident from the calculated percentage error that GGA functional yields better agreement of the lattice parameter with experimental data \cite{2017experimental}. The larger equilibrium lattice constant for studied unit cell is expected from GGA calculations, since during the interaction of valence electron with pseudo core, the later push away the valence electron when GGA functional is used \cite{ggavsldaPhysRevB.51.9521} thereby making the solid softer with large lattice constant and small bulk modulus (as evident from below discussion) than that obtained using LDA. A similar tendency of LDA and GGA have also been observed in previous studies of pernitrides \cite{2014elsevierMANYALI2014706,482014-firstMANYALI2013299}. However, the close agreement between our calculated and experimental results are sufficient enough to estimate the mechanical and vibrational properties of MN$_2$ and understand their behaviour at high pressure. Since the ionic radius increases in the order Si $<$ Ge $<$ Sn, the calculated lattice parameters increase in the order of SiN$_2$ $<$ GeN$_2$ $<$ SnN$_2$  consistent with the recent experimental result \cite{2017experimental}. However, since Niwa \etal\cite{2017experimental} were unable to refine crystallographic parameter for the atomic position of SiN$_2$, there are no available experimental data for N-N and Si-N bond lengths to compare with. It should be noted that N-N bonding in all the structures is greater than that of molecular N$_2$ (1.09 \AA) consistent with the previous studies \cite{2003weihrich2003,2011DING20111357,2017experimental}. The Si-N bond length (1.91 \AA) is slightly greater than the corresponding value of 1.89 \AA\ reported for Si$_3$N$_4$ \cite{si3n4PhysRevB.84.014113} indicating strong covalent bond nature between Si and N atoms. The ratio d$_{M-M}$/d$_{M-N}$ decreases, as we move from lighter to the heavier unit cell. The strong bonding character of M-N and N-N atom in all these compounds partially indicates the hardness of group-14 pernitrides. We studied the effect of pressure on lattice constants and  d$_{M-M}$/d$_{M-N}$ of SiN$_2$, GeN$_2$ and SnN$_2$. The effect of pressure on structural parameters of MN$_2$ is presented in Fig.~\ref{fig:lattice}(a). Lattice parameter decreases uniformly with increasing pressure as shown in Fig.~\ref{fig:lattice}. However, the ratio N-N / M-N increases with pressure indicating M-N bond length decreases faster than N-N, which can be attributed to the strong N-N bonding as compared to M-N bond. We have shown the results for GeN$_2$ in Fig.~\ref{fig:lattice}(b), and similar results hold for SiN$_2$ and SnN$_2$ and shown in Fig.~S1 of supporting information. An increase in pressure from 0 to 30 GPa cause decrease in  Si-N, Ge-N, and Sn-N bond length of  SiN$_2$, GeN$_2$ and SnN$_2$ by 2.73 \%, 3.6 \% and 4.29 \% respectively, whereas N-N bond length decreases by 1.7 \%, 1.48 \%, and 1.63 \%, respectively.

\subsection{Mechanical Properties}
\begin{figure}
\centering
\includegraphics[width=0.5\textwidth]{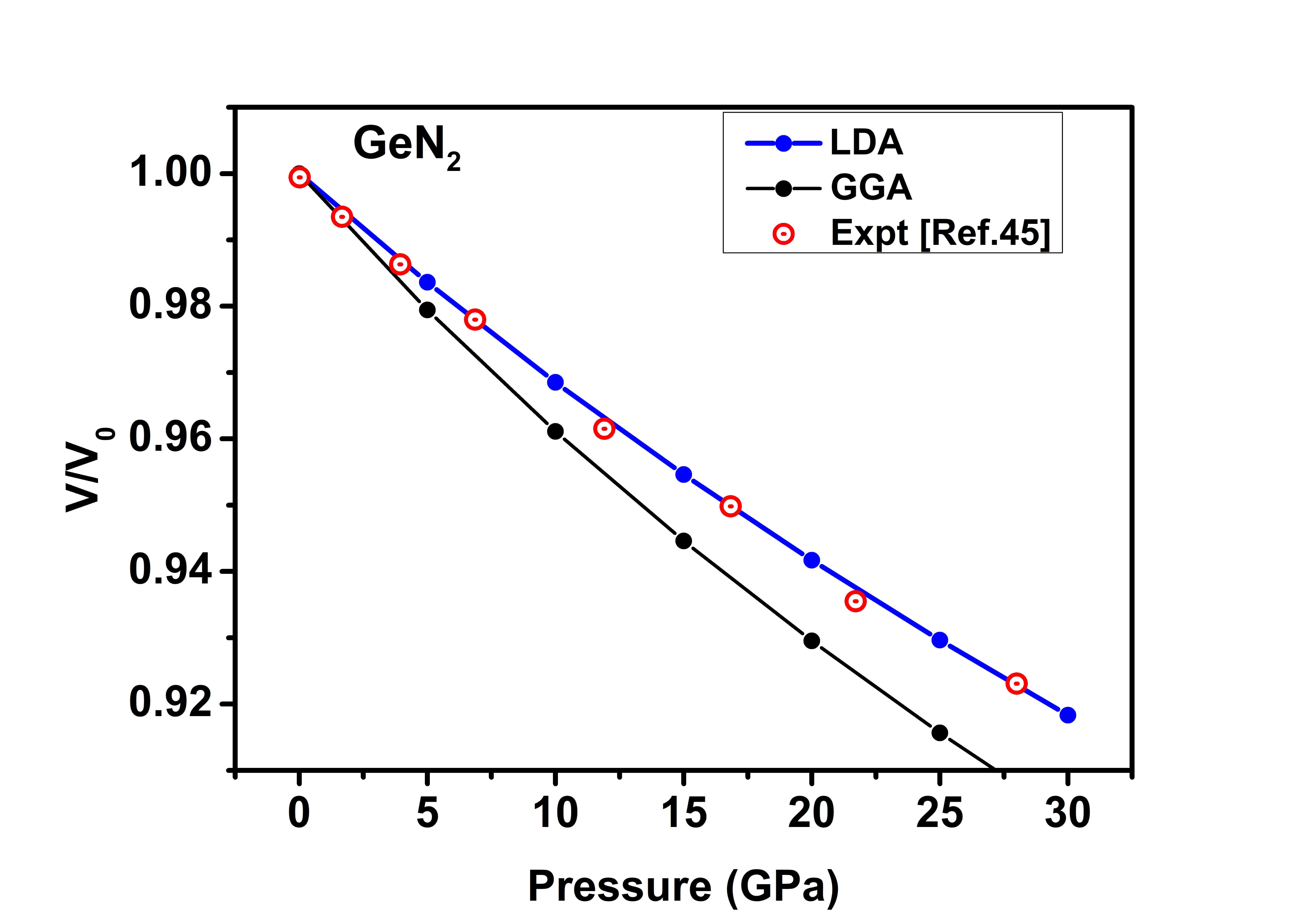}
\caption{Compressibility of GeN$_2$ calculated using LDA and GGA functional and compared with experimental data of Niwa et. al. \cite{2017experimental}}
\label{fig:lattice1}
\end{figure}

\begin{figure}
\centering
\includegraphics[width=0.6\textwidth]{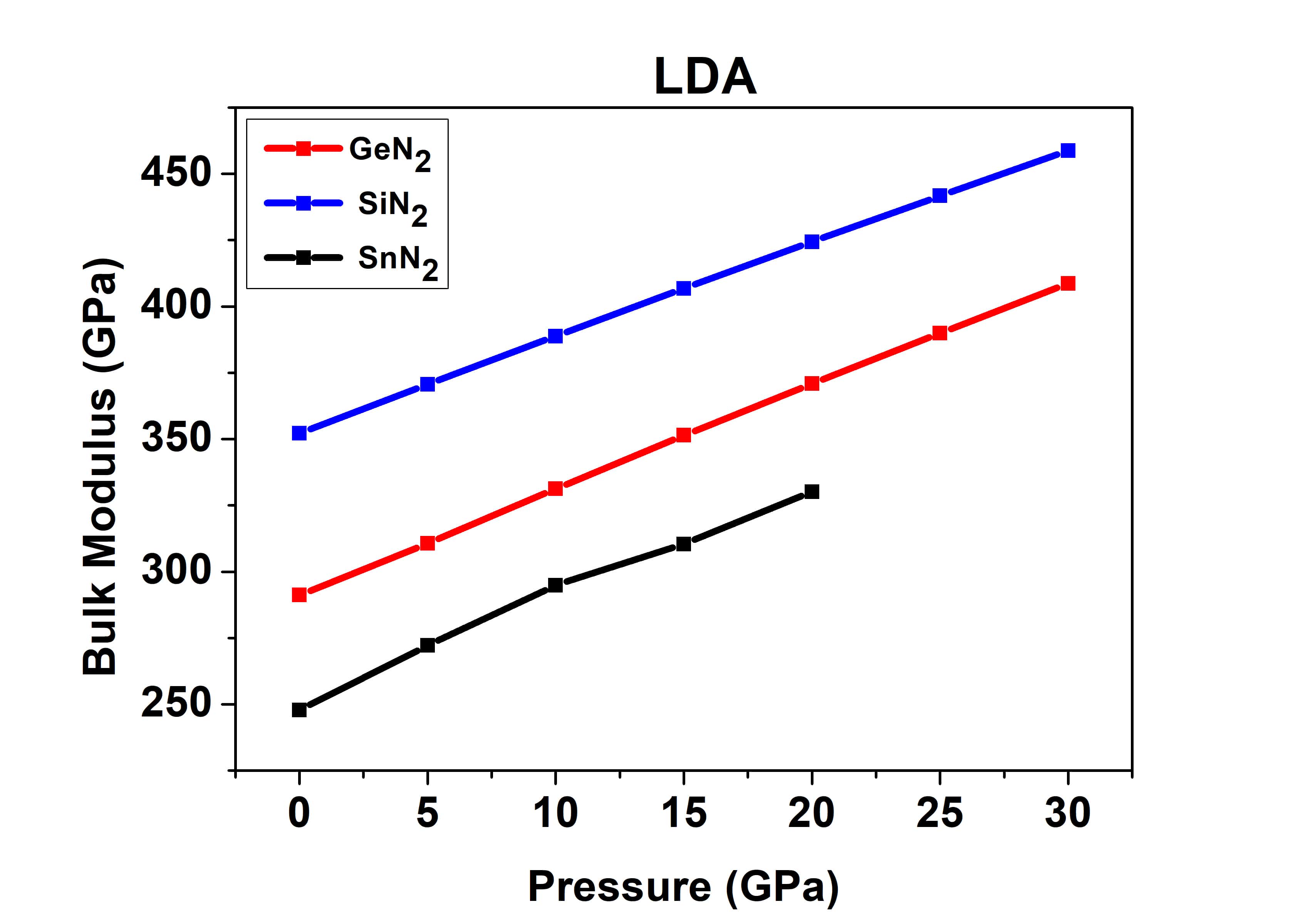}
\includegraphics[width=0.6\textwidth]{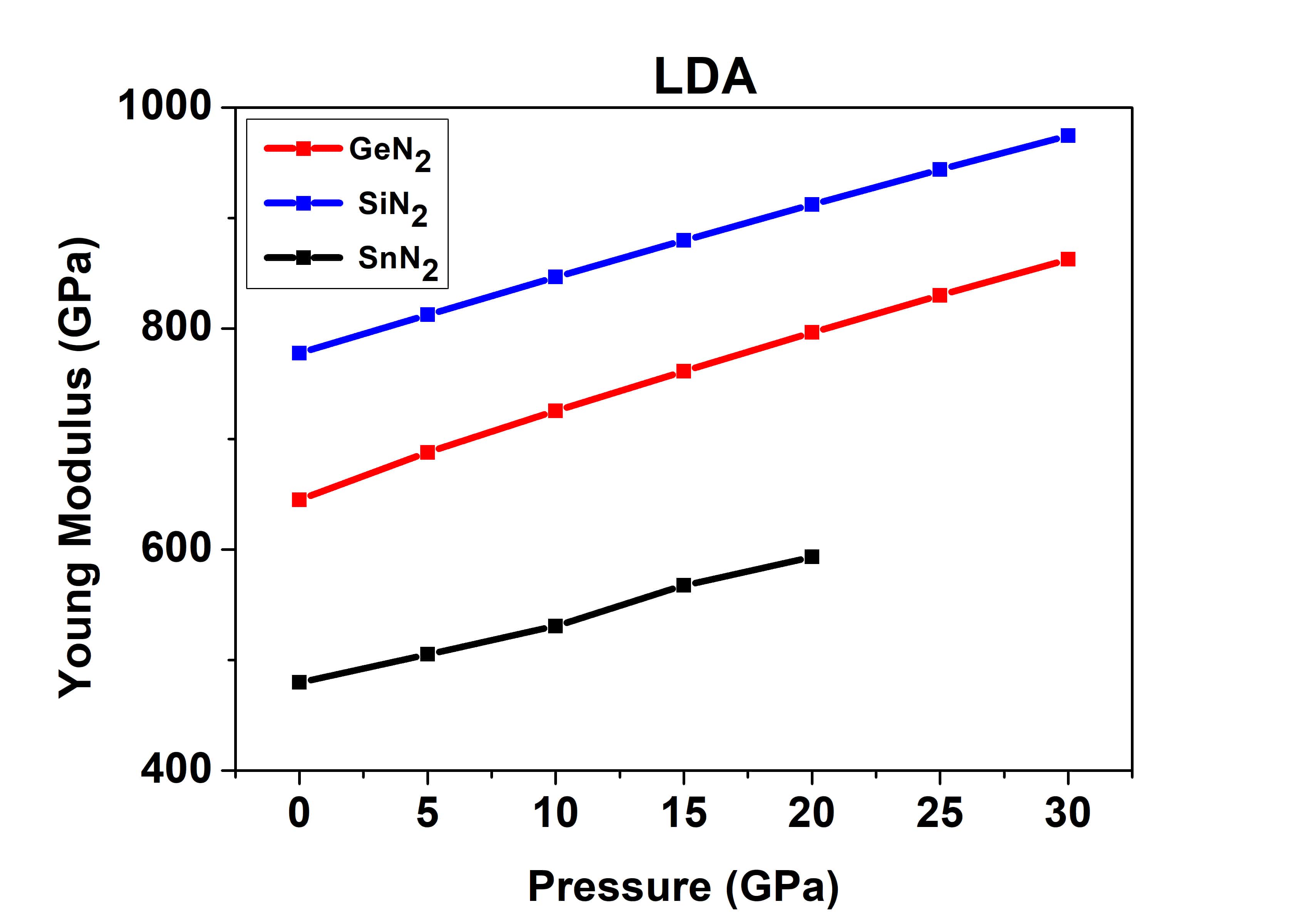}
\includegraphics[width=0.6\textwidth]{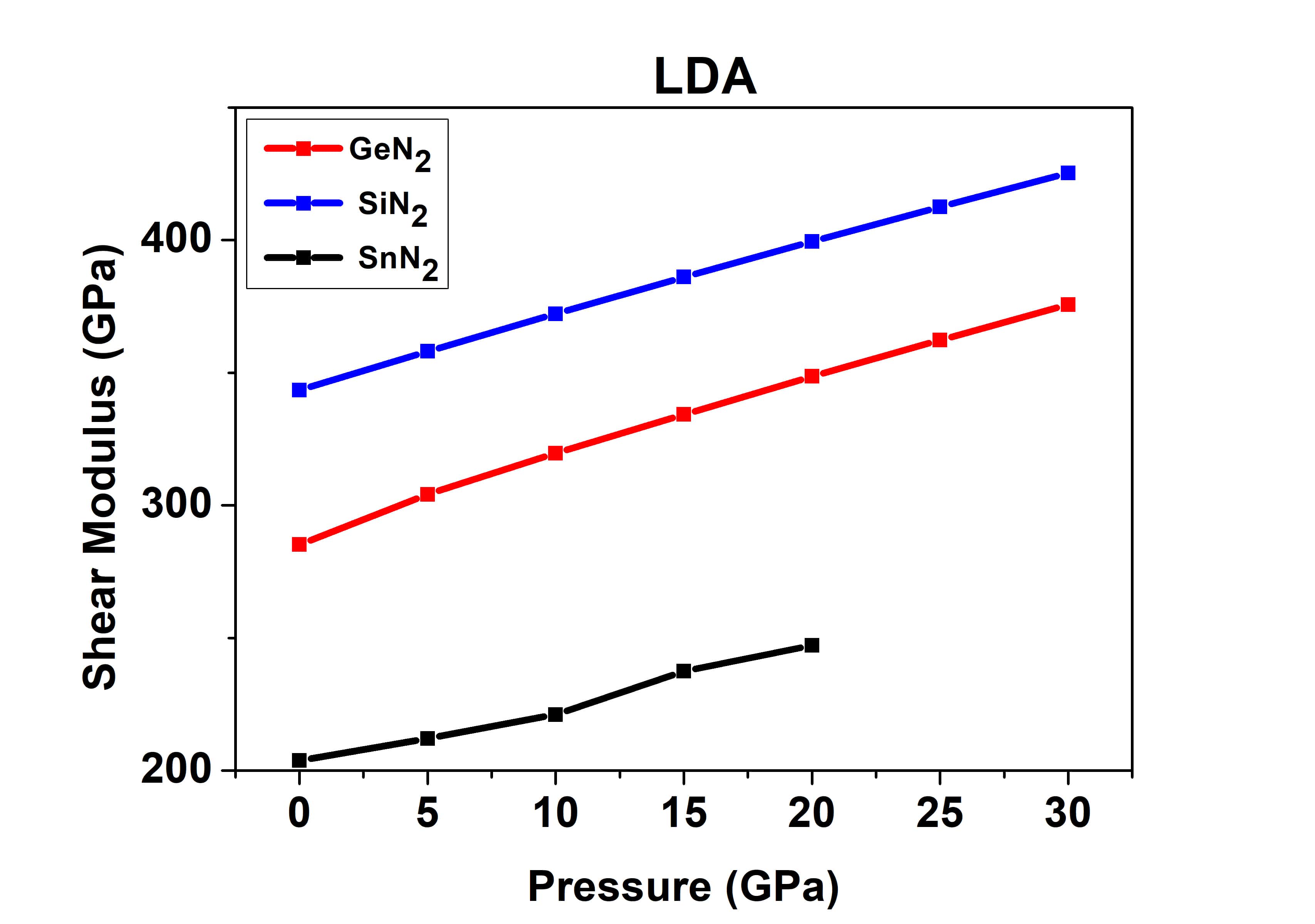}
\caption{Bulk modulus (B), Shear modulus (G) and Young Modulus (Y) of GeN$_2$,  SiN$_2$ and  SnN$_2$ at different pressures using LDA.}
\label{fig:elastic}
\end{figure}

Investigating mechanical properties of MN$_2$ are done by calculating the second order elastic constants, which are important in the industrial applications and also regarding as mechanical stability criteria. Prior to evaluation of elastic constants for determining their mechanical properties, we test the validity of two functionals in deforming the crystal. Fig.~\ref{fig:lattice1} shows the change in the volume of GeN$_2$ with pressure. It is calculated using both LDA and GGA and compared with the experimental data \cite{2017experimental}. Similar results for SiN$_2$ and SnN$_2$ are shown in Fig.~S2 of supporting information. The compression behavior of MN$_2$ calculated using LDA follows the experimental observations, whereas, the GGA underestimates the unit cell volume at high pressure. The calculated values of elastic constants and moduli for MN$_2$ are listed in Tab.~\ref{tab:example_2}. Comparison of zero pressure elastic constants calculated for MN$_2$ shows a significant difference between LDA and GGA values. Similar discrepancies in elastic constant have also been reported in previous studies of pyrite dinitride \cite{2011DING20111357}. Our calculated bulk modulus of MN$_2$ at zero pressure using LDA is close to the experimental values \cite{2017experimental}, whereas, much lower value is obtained using GGA, which can be attributed to the over-binding of lattice constants by GGA. However, the suitability of LDA over GGA in describing mechanical properties including material's mechanical stability has been emphasized in pyrite CN$_2$ \cite{2011DING20111357}. Further, with our investigation on MN$_2$, one can realize that the GGA is not good for determining mechanical properties of pernitride of group-14 elements. The mechanical stability of pyrite MN$_2$ compounds was determined by checking the Born stability condition \cite{bookbornFedorov2012}. For cubic crystals, the stability criteria reads as:
C$_{11}$ $>$ 0, C$_{44}$ $>$ 0, C$_{11}$  $>$  $|$C$_{12}$$|$ , C$_{11}$ + 2 $|$C$_{12}$$|$  $>$ 0, C$_{12}$ $<$ B < C$_{11}$. The elastic constants of MN$_2$ calculated using LDA were shown to obey the stability criteria. From Tab.~\ref{tab:example_2}, it can be seen that the bulk modulus decreases as from SiN$_2$ to SnN$_2$, implying that the compressibility of pyrite pernitride increases down the group with CN$_2$ being least compressible, and SnN$_2$ being the most compressible among MN$_2$, which can be attributed to the increasing bond strength of M-N bond length. The same trend was also observed on group-14 pernitride with orthorhombic symmetry \cite{2014elsevierMANYALI2014706}. Shorter M-N bond length is more incompressible, whereas, pernitride with large M-N bond length are easily compressible due to its weak strength. Shear modulus determines the materials' response to change in shape at fixed volume. SiN$_2$ has greater resistance to change in shape than GeN$_2$ and SnN$_2$. The stability of a crystal against shear stress is evaluated from its Poisson's ratio. MN$_2$ are brittle, since they have poisons ratio less than 0.33. The small value of Poisson's ratio further indicates a high degree of covalent bonding in these nitrides. Similar to the behaviour of bulk and shear modulus; Young's modulus obtained as a combination of both bulk and shear modulus, decreases down the group of their corresponding pernitride as seen from Tab.~\ref{tab:example_2}. For a material to be hard, it should have high bulk modulus to withstand the volume decrease caused by the applied pressure and a high shear modulus or low Poisson's ratio in order to resist the deformation caused by the applied load from different directions. SiN$_2$ has a bulk modulus of 352.17 GPa, which is less than the known hard nitride materials such as cubic-BN (400 GPa) \cite{bn757} and CN$_2$ (405 GPa) \cite{2003weihrich2003}, whereas, higher than GeN$_2$ and SiN$_2$. Poisson's ratio of SiN$_2$ is comparable with GeN$_2$, whereas, much lower than SnN$_2$. The comparable bulk modulus of SiN$_2$ with cubic BN and CN$_2$ \cite{2003weihrich2003} shows that SiN$_2$ can be regarded as a hard material. However, since both bulk modulus and shear modulus are less than that of diamond \cite{diamondMANYALI2013299}, none of group-14 elemental nitrides is harder than diamond at 0 GPa. Hardness property of materials straight away demands high pressure studies for investigating the modification of elastic properties under pressure. The effect of pressure on elastic constants, bulk modulus, Young's modulus and shear modulus of MN$_2$ is shown in Fig.~\ref{fig:elastic}. All elastic moduli of MN$_2$ increases with pressure thereby increasing their hardness in the considered pressure range.

\subsection{Vibrational Properties}

\begin{figure}
\centering
\includegraphics[width=0.5\textwidth]{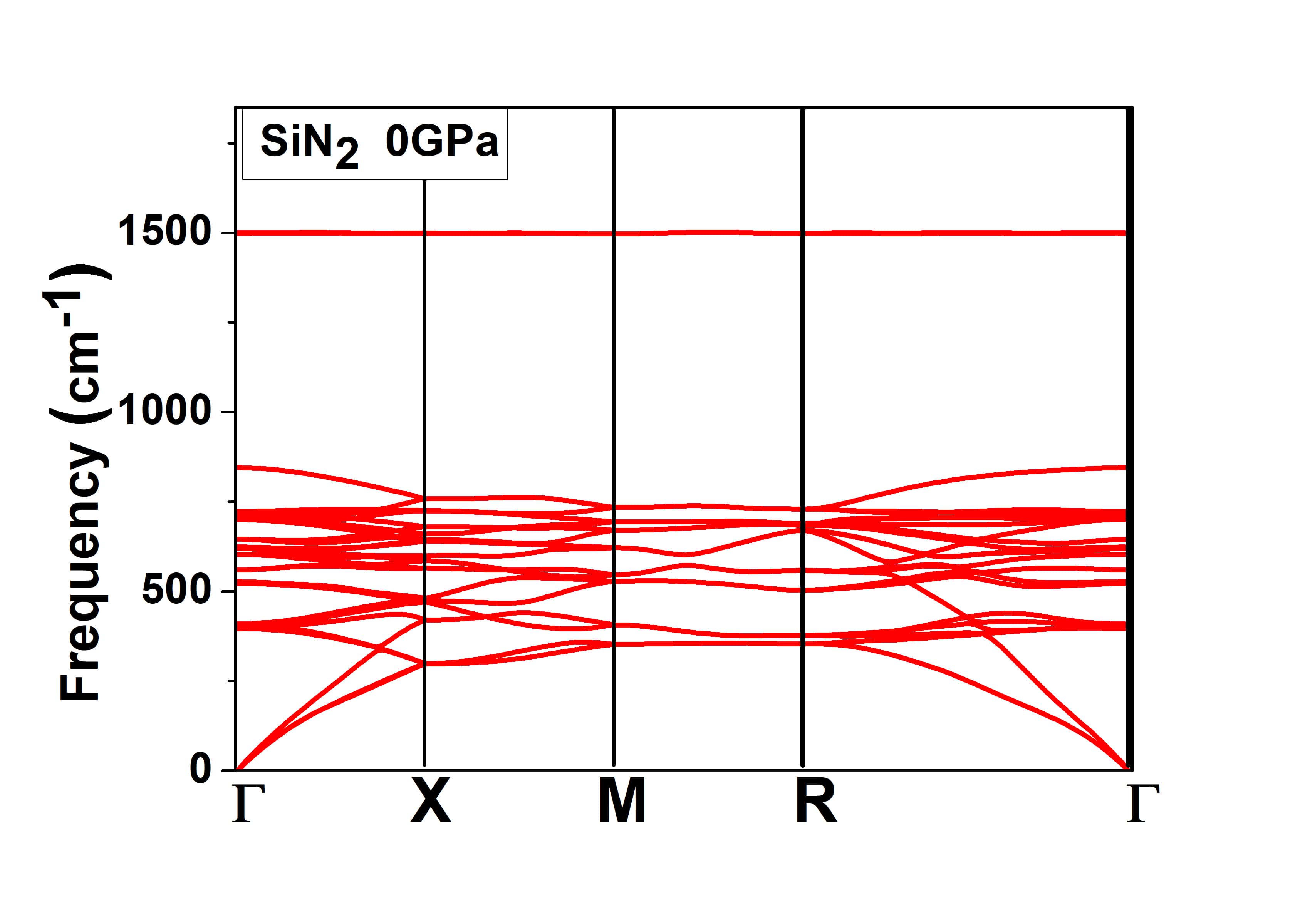}
\includegraphics[width=0.5\textwidth]{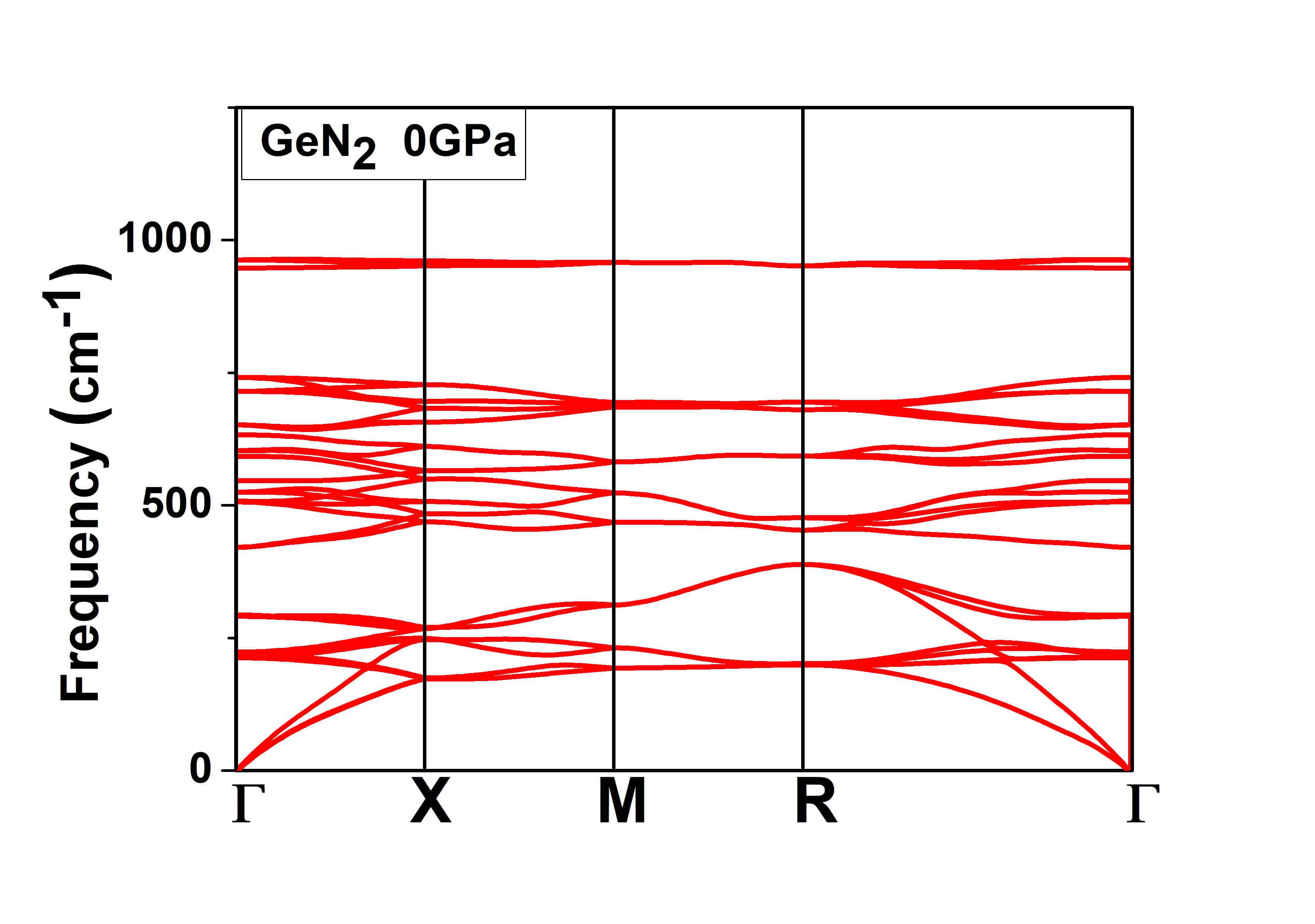}
\includegraphics[width=0.5\textwidth]{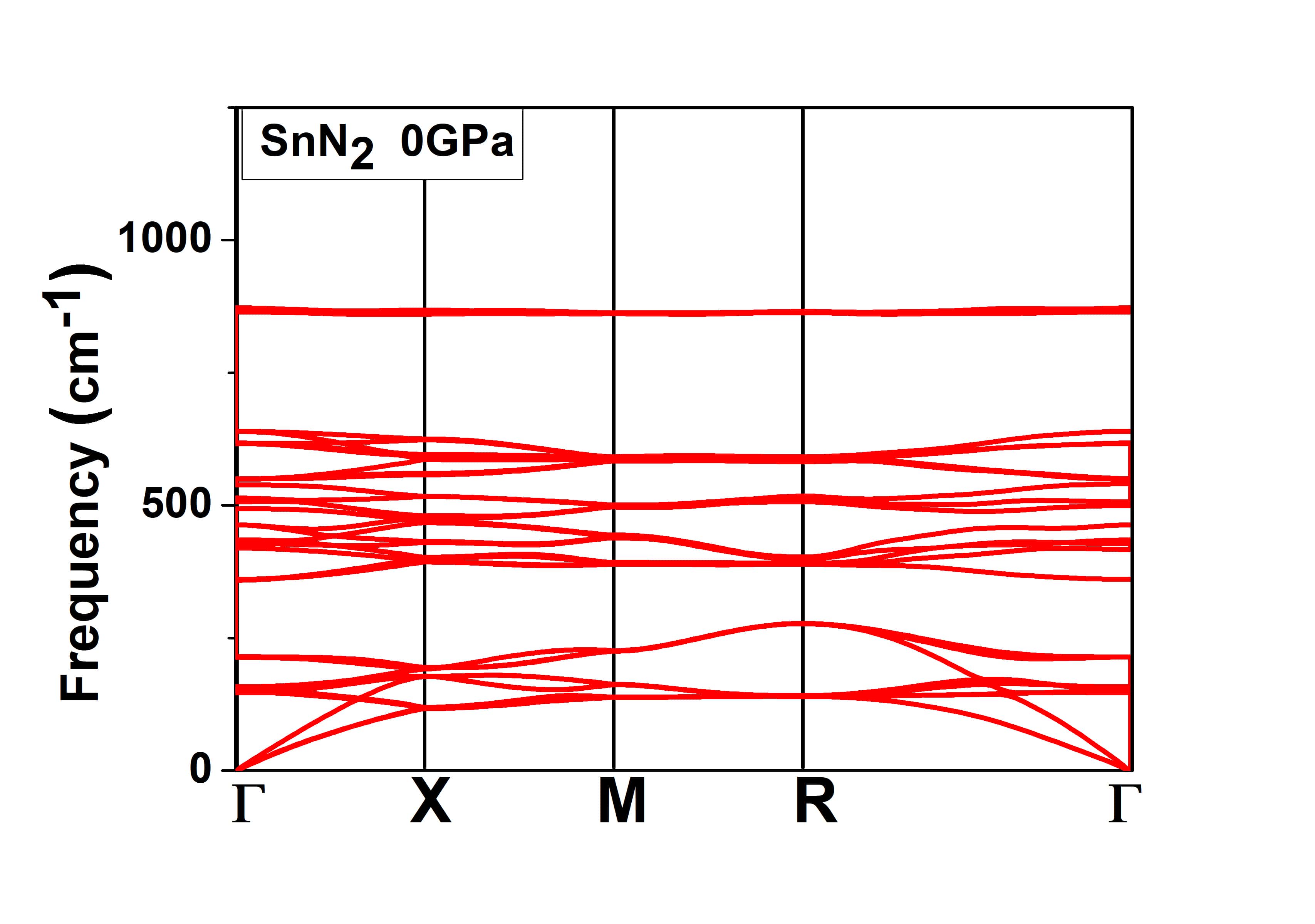}
\caption{Phonon dispersion curves of SiN$_2$, GeN$_2$ and SnN$_2$ using GGA at 0 GPa.}
\label{fig:phonon}
\end{figure}

\begin{figure}
\centering
\includegraphics[width=0.6\textwidth]{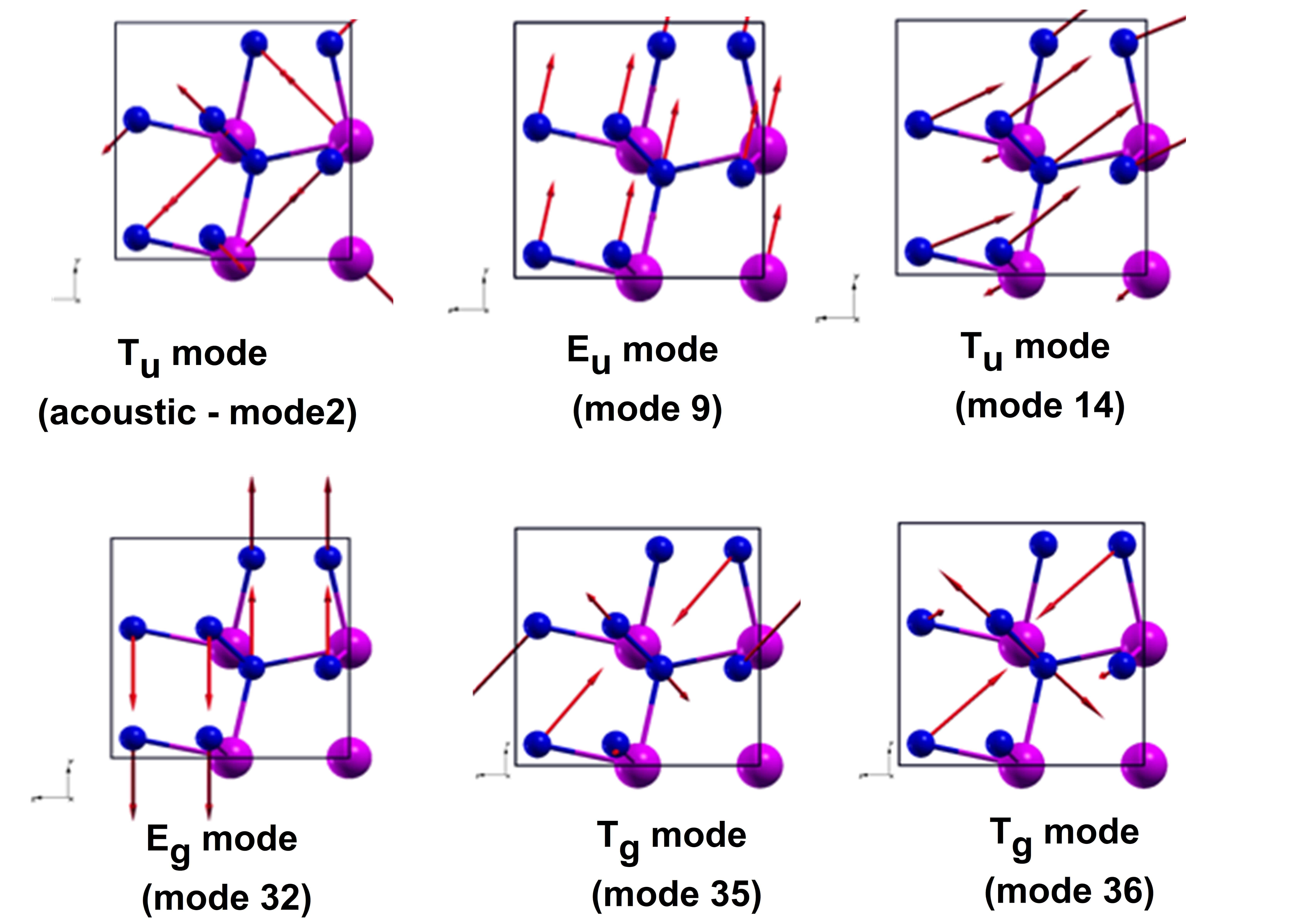}
\caption{Eigen vectors of low, mid and high frequency phonon modes of SiN$_2$, GeN$_2$ and SnN$_2$. Violet and Blue sphere represents M and N atom respectively.}
\label{fig:modes}
\end{figure}

\begin{figure}
\centering
\includegraphics[width=0.4\textwidth]{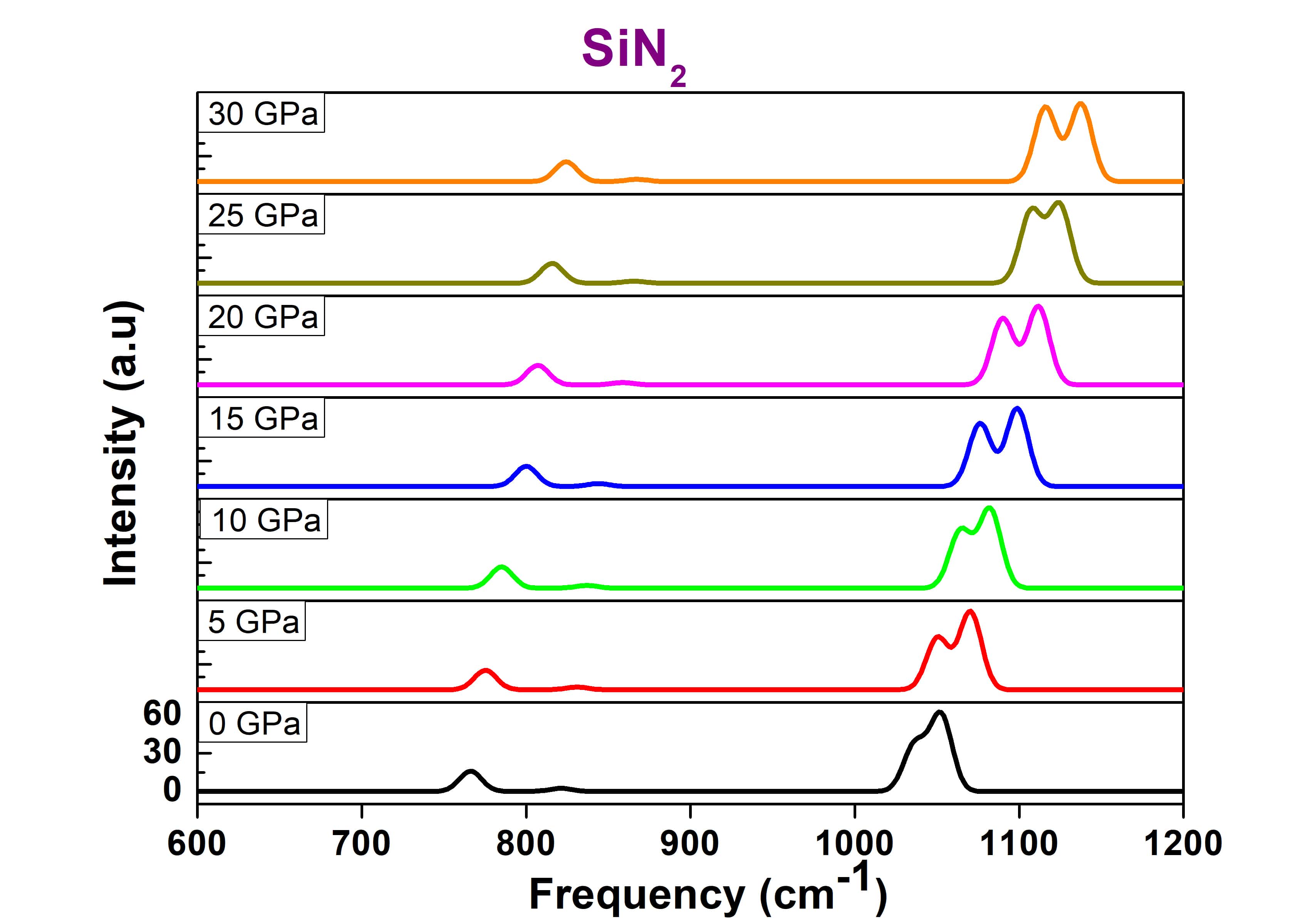}
\includegraphics[width=0.4\textwidth]{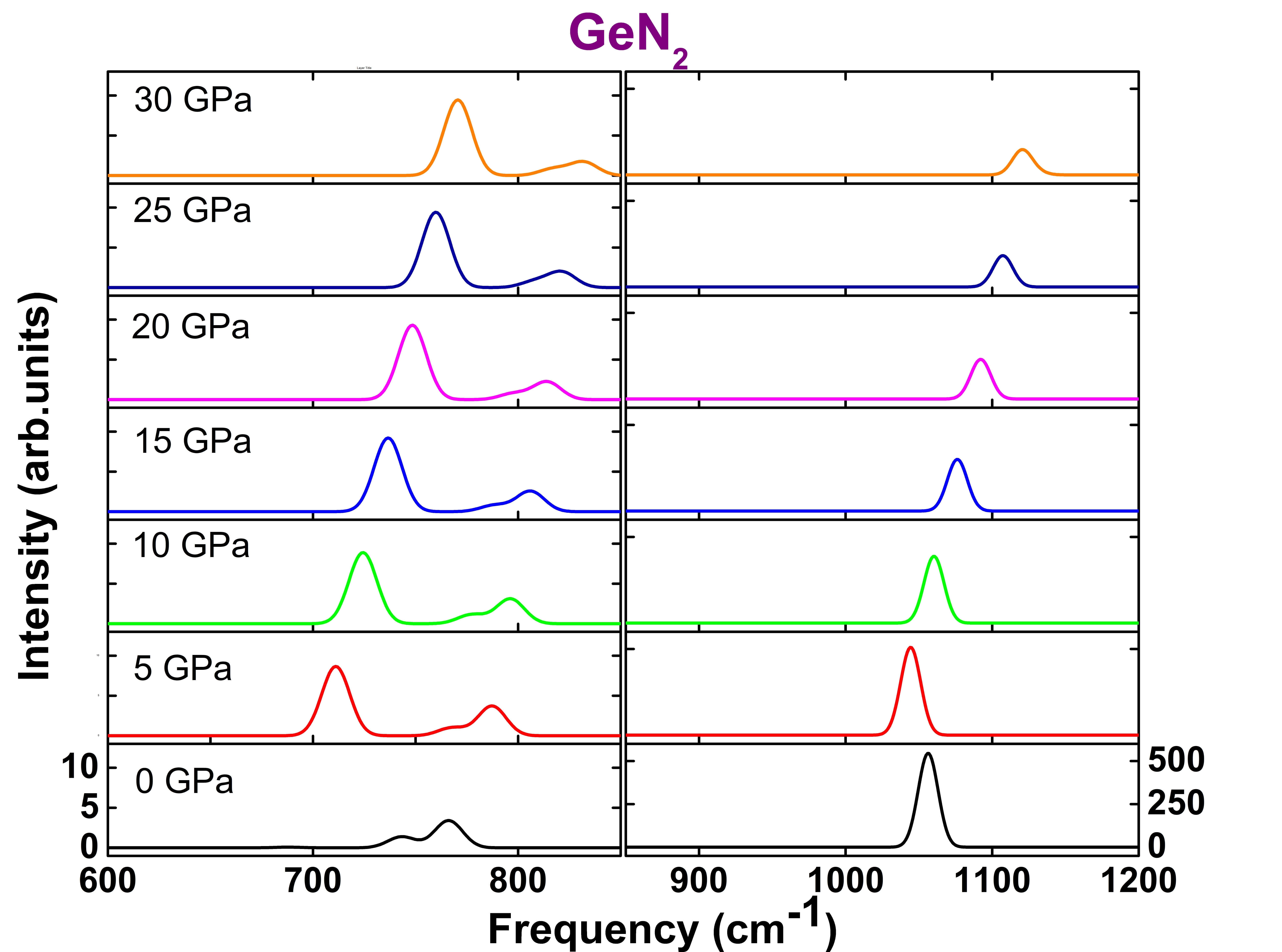}
\includegraphics[width=0.4\textwidth]{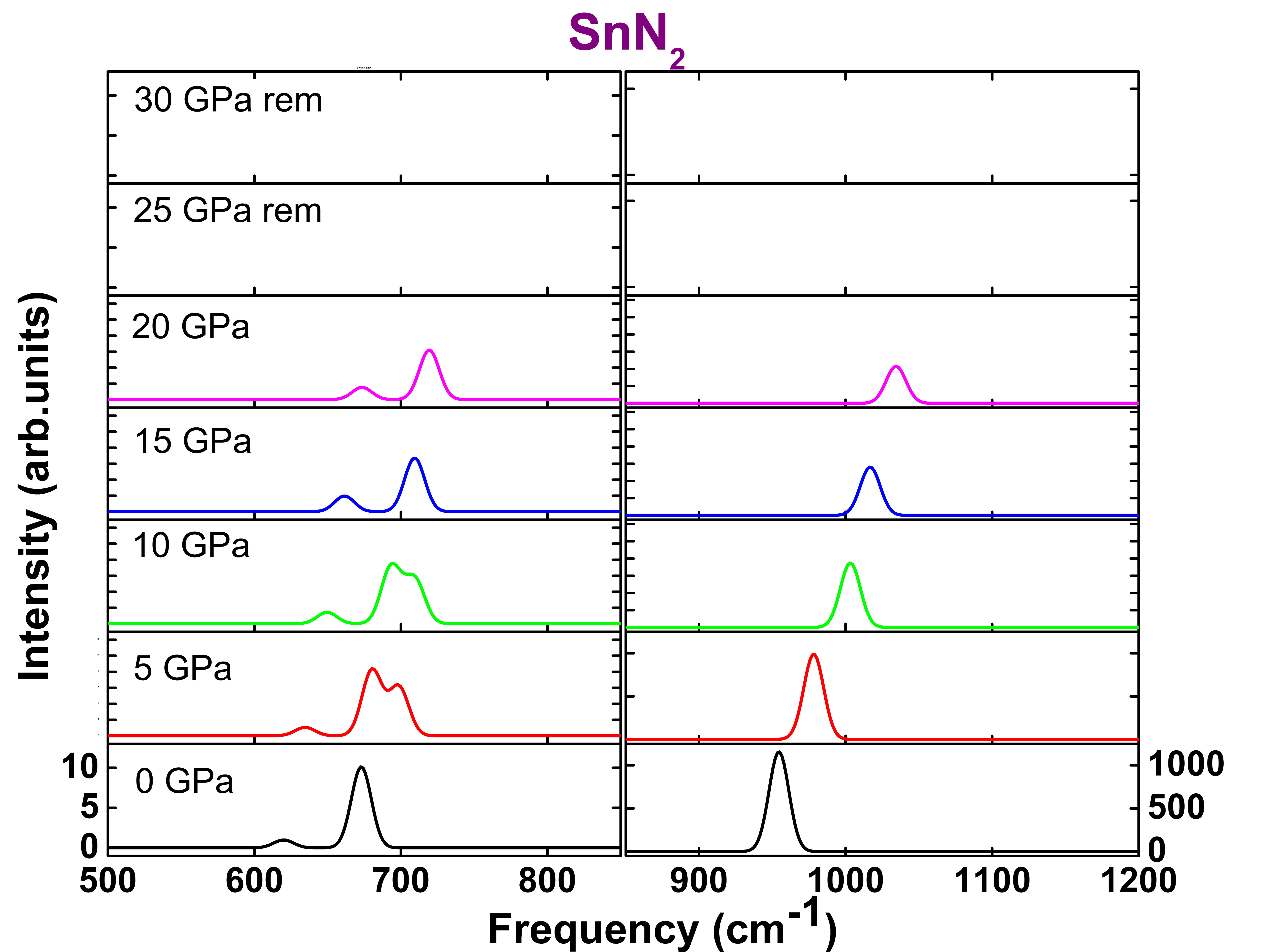}
\caption{Calculated Raman spectra of SiN$_2$, GeN$_2$ and SnN$_2$ at different pressures.}
\label{fig:Raman}
\end{figure}

Being nonmetals, MN$_2$ can be easily characterized by their vibrational properties. To confirm the dynamical stability of our studied structures, we calculated phonon dispersion curves of MN$_2$ throughout the Brillouin zone. The absence of imaginary frequencies in all phonon dispersion curves shows the dynamical stability of considered compounds. The phonon dispersion curves of MN$_2$ consists of 36 phonon branches corresponding to 12 atoms in the unit cell. The calculated phonon dispersion curves along $\Gamma$ - X - M - R - $\Gamma$ symmetry directions are presented in Fig.~\ref{fig:phonon}. The phonon frequencies of MN$_2$ can be grouped into three categories: low, mid and high frequency modes. Few selected eigen vectors from these three categories are shown in Fig.~\ref{fig:modes}. The eigen vector analysis shows that the low frequency optical modes are due to coupled vibrations of M (M = Si, Ge and Sn) and N atoms. As the frequency increases, the contribution of M atoms decreases as shown in Fig.~\ref{fig:modes}. The high frequency regions are mainly due to the vibration of nitrogen atoms, which includes the stretching vibrations of nitrogen dimers. We further calculated the Raman spectrum using first-principles simulations, and compared with the experimental measurements and revealed the unobserved Raman bands in the experimental measurements. The allowed Raman active modes are $\Gamma$ = A$_g$ + E$_g$ + 3T$_g$. Fig.~\ref{fig:Raman}  presents the Raman spectra of MN$_2$ at different pressure. We observed more than one distinct Raman peaks for all the nitrides, which were not observed in the recent experimental study \cite{2017experimental}. We propose that the relative intensity of low frequency E$_g$ modes are too low, hence these peaks might got suppressed in the experimental study. The state of nitrogen dimer can be inferred from the Raman spectrum. The N - N, N = N and N $\equiv$ N have their stretching vibration modes in the range of 800 - 1000, 1300 - 1500 and 2300 - 2400 cm$^{-1}$, respectively, and bond length of ~ 1.4, 1.2 - 1.3 and 1.1 \AA\, respectively. The stretching frequency and along with the calculated N - N bond length (Tab.~\ref{tab:example_1}) indicates the single bond character of N$_2$ dimer at 0 GPa. At higher pressures, A$_g$ and T$_g$ modes of SiN$_2$ hardens in such a way that their frequency difference is noticeable and will appear distinctly at high pressure Raman measurements.

\subsection{Electronic Properties}

\begin{figure}
\centering
\includegraphics[width=0.6\textwidth]{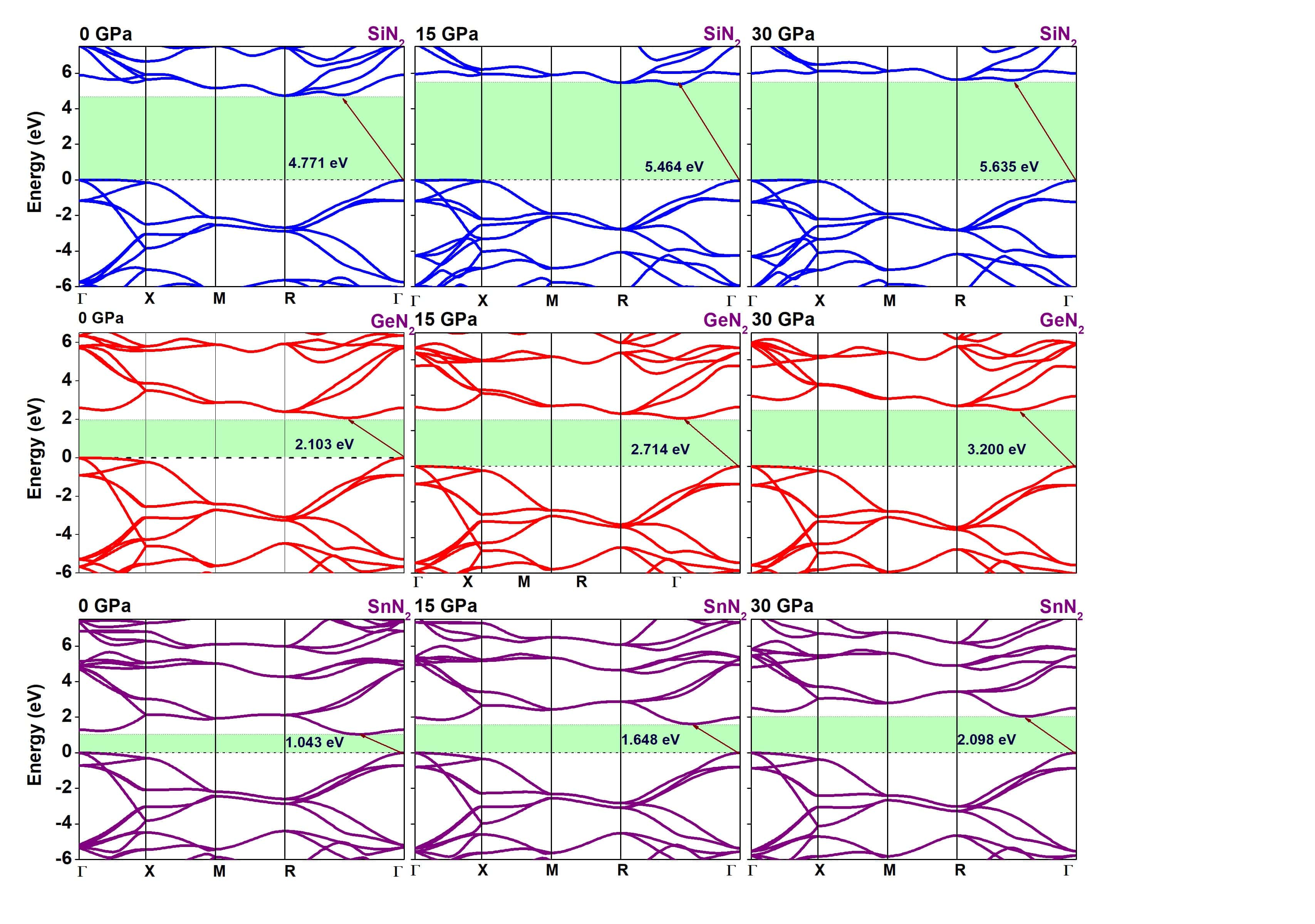}
\caption{Calculated electronic band structure of of  SiN$_2$, GeN$_2$ and SnN$_2$ using GGA at different pressures.}
\label{fig:bandstructure}
\end{figure}

 To give insight into the electronic properties of group-14 nitrides, we calculated the electronic band structure of MN$_2$  at ambient and high pressure conditions. The electronic band gap of pyrite SnN$_2$ is unknown to the best of our knowledge. Though DFT calculations fail to produce accurate band gap values owing to challenges in describing the exchange and correlation potential, it won't change the qualitative analysis of electronic behaviour of MN$_2$ under pressure. The electronic structure calculations show that the MN$_2$ exhibit indirect band gap with valence band maximum (VBM) at the Brillouin zone center and conduction band minimum (CBM) lying along $\Gamma$ - R direction. We found that SiN$_2$ is a wide band gap (4.77 eV) insulator, which is consistent with earlier studies \cite{2014rscC4RA11327F}, whereas, GeN$_2$ and SnN$_2$ are indirect band gap semiconductors. The calculated electronic band gaps of MN$_2$ are presented in Tab.~\ref{tab:example_3} along with the previously reported results \cite{2003weihrich2003,2014rscC4RA11327F}. The electronic band gap of pyrite MN$_2$ follows the order SiN$_2$ $>$ GeN$_2$ $>$ SnN$_2$ and this trend can be attributed to increasing bond distance (Fig.~\ref{fig:lattice}(b)), which results in lower orbital overlap, thereby increasing the band gap. The calculated band gap of pyrite SiN$_2$ and GeN$_2$ are lower than that of proposed orthorhombic SiN$_2$ and GeN$_2$. Further, we studied the electronic band gap of these compounds at high pressures as these pernitrides are foreseen in high-pressure device applications. Electronic band structure of MN$_2$ at different pressures are presented in Fig.~\ref{fig:bandstructure}. When subjected to external pressure, the electronic band gap of MN$_2$ increases with pressure, however, no indirect-direct band gap transition is observed in the considered pressure range nor any significant shift is observed in the position of VBM and CBM.

 \subsection{Conclusions}

In this work, we explored the structural, electronic, mechanical and vibrational behaviour of recently synthesized SiN$_2$, GeN$_2$ and SnN$_2$ at ambient and high pressure conditions. To provide better estimates for all studied properties, we have performed the structural calculations at ambient and high pressure condition using both LDA and GGA. At ambient condition, the structural parameters calculated using GGA agrees well with the experimental data \cite{2017experimental}. However, when checked for the high pressure compressibility behaviour of considered systems, LDA yields identical behavior to that of obtained in the experiments. Hence, the high pressure mechanical properties of SiN$_2$, GeN$_2$ and SnN$_2$ were calculated using LDA. The elastic moduli calculated using LDA at 0 GPa gives better estimate of values than GGA, confirming the reliability of GGA over LDA. The calculated elastic parameters showed that SiN$_2$ is a hard material among considered pernitrides and elastic moduli increases with pressure for SiN$_2$, GeN$_2$ and SnN$_2$. We further predicted the electronic band gap of SiN$_2$, GeN$_2$ and SnN$_2$ at high pressures. The insulating nature of SiN$_2$ and semiconducting nature of GeN$_2$ and SnN$_2$ do not change with pressure. However, the electronic band gap increases with pressure. The lattice dynamical stability of our considered structures were confirmed by phonon dispersion calculations and show no imaginary frequencies throughout the brillouin zone. The high pressure structures of SiN$_2$, GeN$_2$ and SnN$_2$ were further characterized using Raman spectral analysis.

\section{Acknowledgements}
Authors acknowledge Ministry of Earth Sciences, Govt. of India for providing financial assistance under Active Fault Mapping programme (MoES/P.O.(seismo)/1(270)/AFM/2015).

\section*{References}


\newpage

\begin{table*}
        \centering
    \begin{tabular}{  l  p{1.1cm}  p{1.1cm}  p{1.1cm} p{1.1cm}  p{1.1cm}  p{1.1cm} p{1.1cm}  p{1.1cm}  p{1.1cm}}
        \hline
        \hline
         & \multicolumn{3}{c}{SiN$_{2}$} & \multicolumn{3}{c}{GeN$_{2}$} & \multicolumn{3}{c}{SnN$_{2}$}  \\
         & LDA &GGA &Others&LDA &GGA&Others&LDA&GGA&Others \\
        \hline
        \textbf{a} (\AA) & 4.394 (0.97) & 4.466 (0.65) & 4.437$^a$  4.405$^b$ 4.391$^c$ 4.420$^d$ 4.455$^e$ &           4.584 (3.64)& 4.730 (1.28)& 4.670$^a$ 4.596$^b$ 4.579$^c$ 4.647$^d$ & 4.904 (3.2)&5.117 (0.97)&5.068$^a$\\
        \hline
        \textbf{N-N} (\AA) &1.442&1.468&1.454$^d$&1.423 (3.59)&1.443 (2.23)&1.476$^a$ 1.428$^d$ &1.452&1.468&1.516$^a$ \\
        \hline
        \textbf{A-N} (\AA) & 1.875&1.905&1.886$^d$&1.97 (1.55)&2.036 (1.75)& 2.001$^a$ 1.998$^d$ &2.118 (2.9)&2.217 (1.62)& 2.182$^a$\\

        \hline
        \hline
    \end{tabular}
        \caption{Calculated structural parameters of SiN$_{2}$, GeN$_{2}$ and SnN$_{2}$ using LDA and GGA (percentage error with respect to experimental values are in parenthesis).}
    \label{tab:example_1}
    \begin{tabbing}
$^{\mathrm{a}}$ Expt.\cite{2017experimental}
\hspace{5pt} \=
$^{\mathrm{b}}$GGA result of {Ref.~\cite{2011DING20111357}}
\hspace{5pt} \=
$^{\mathrm{c}}$ LDA result of {Ref.~\cite{2011DING20111357}}
\hspace{5pt} \=
$^{\mathrm{d}}$ {Ref.~\cite{2003weihrich2003}}
\hspace{5pt} \=
$^{\mathrm{e}}$ {Ref.~\cite{2014rscC4RA11327F}}
\hspace{5pt} \=
\end{tabbing}
\end{table*}

\begin{table*}
	\centering
      \begin{tabular}{  l  p{1.1cm}  p{1.1cm}  p{1.1cm} p{1.1cm}  p{1.1cm}  p{1.1cm} p{1.1cm}  p{1.1cm}  p{1.1cm}}
    	\hline
        \hline
         & \multicolumn{3}{c}{SiN$_{2}$} & \multicolumn{3}{c}{GeN$_{2}$} & \multicolumn{3}{c}{SnN$_{2}$}  \\
         		  & LDA & GGA & Others  & LDA & GGA & Others & LDA & GGA & Others \\
        \hline
        \textbf{C$_1$$_1$}  & 783.2&696.8&783.74$^a$ 801.00$^b$ 704.00$^c$ &588.1&455.9&468.54$^a$ 513.69$^b$&490.6&511.2\\
        \hline
       \textbf{C$_1$$_2$}  & 136.6&128.0&408.58$^a$ 534.32$^b$ 127.00$^c$&142.7&118.6&87.57$^a$ 90.81$^b$&126.4&-72.5\\
        \hline
        \textbf{C$_1$$_4$}  & 357.6&339.4&783.74$^a$ 408.75$^b$ 379.00$^c$&336.6&287.9&329.5$^a$ 330.25$^b$&219.7&174.1\\
        \hline
         \textbf{B}  & 352.17&317.62&354.00$^d$ 356.94$^a$ 356.88$^b$ 320.00$^c$ &291.15&231.03&284.00$^d$ 214.56$^a$ 231.77$^b$&247.82&122.03&219$^d$]\\
        \hline
        \textbf{Y}  & 777.61&712.21&739.30$^a$ 762.15$^b$&645.12&521.93&440.96$^a$ 486.40$^b$&479.85&405.57\\
        \hline
         \textbf{G}  & 343.471&316.18&370.55$^a$ 376.60$^b$ 343.00$^c$&285.27&232.33&264.47$^a$ 276.18$^b$&203.80&214.4\\
        \hline
        \textbf{P} &   0.13&0.13&0.155$^a$ 0.144$^b$ 0.1504$^c$&0.12 &0.12&0.158$^a$ 0.150$^b$&0.18&-0.05\\\\
        \hline
        \hline
    \end{tabular}
	\caption{Calculated elastic constants (C$_{11}$, C$_{12}$ and C$_{14}$), Bulk modulus (B), Young modulus (Y), Shear modulus (G) and Poisson's ration (P) of SiN$_{2}$, GeN$_{2}$ and SnN$_{2}$ along with previously reported results.}
    \label{tab:example_2}
\begin{tabbing}
    $^{\mathrm{a}}$ GGA result of {Ref.~\cite{2011DING20111357}}
    \hspace{5pt} \=
$^{\mathrm{b}}$ LDA result of {Ref.~\cite{2011DING20111357}}
\hspace{5pt} \=
$^{\mathrm{c}}$ {Ref.~\cite{2014rscC4RA11327F}}
\hspace{5pt} \=
$^{\mathrm{d}}$ Expt.\cite{2017experimental}
\hspace{5pt} \=
\end{tabbing}
\end{table*}

\begin{table}
	\centering
    \begin{tabular}{  l  p{1.1cm}  p{1.1cm}  p{1.1cm} p{1.1cm}  p{1.1cm}  p{1.1cm} }
    	\hline
        \hline
         & \multicolumn{2}{c}{SiN$_{2}$} & \multicolumn{2}{c}{GeN$_{2}$} & \multicolumn{2}{c}{SnN$_{2}$}  \\
         		  & GGA & Others  & GGA & Others & GGA & Others \\
        \hline
        \textbf{E$_{g}$} (eV) &4.77&5.50\cite{2003weihrich2003} 5.13\cite{2014rscC4RA11327F} & 2.10&1.40\cite{2003weihrich2003} &1.40\\
        \hline
        \hline
    \end{tabular}
	\caption{Calculated electronic bandgap of SiN$_{2}$, GeN$_{2}$ and SnN$_{2}$ using GGA.}
    \label{tab:example_3}
\end{table}
\newpage

\end{document}


\bibliographystyle{agsm} 
\title[First-Principles High Pressure Studies on Group-14 Element Pernitrides]{First-Principles High Pressure Studies on Group-14 Element Pernitrides}

\author{Sharad Babu Pillai$^1$, Himadri R. Soni$^{2*}$ and Prafulla K. Jha$^1$}

\address{$^1$ Department of Physics, Faculty of Science, The M. S. University of Baroda, Vadodara-390002, India}
\address{$^2$ School of Sciences, Indrashil University, Rajpur, Kadi 382740, India}
\ead{*himadri.iu@gmail.com}

\begin{figure}[H]
\centering
\includegraphics[width=0.5\textwidth]{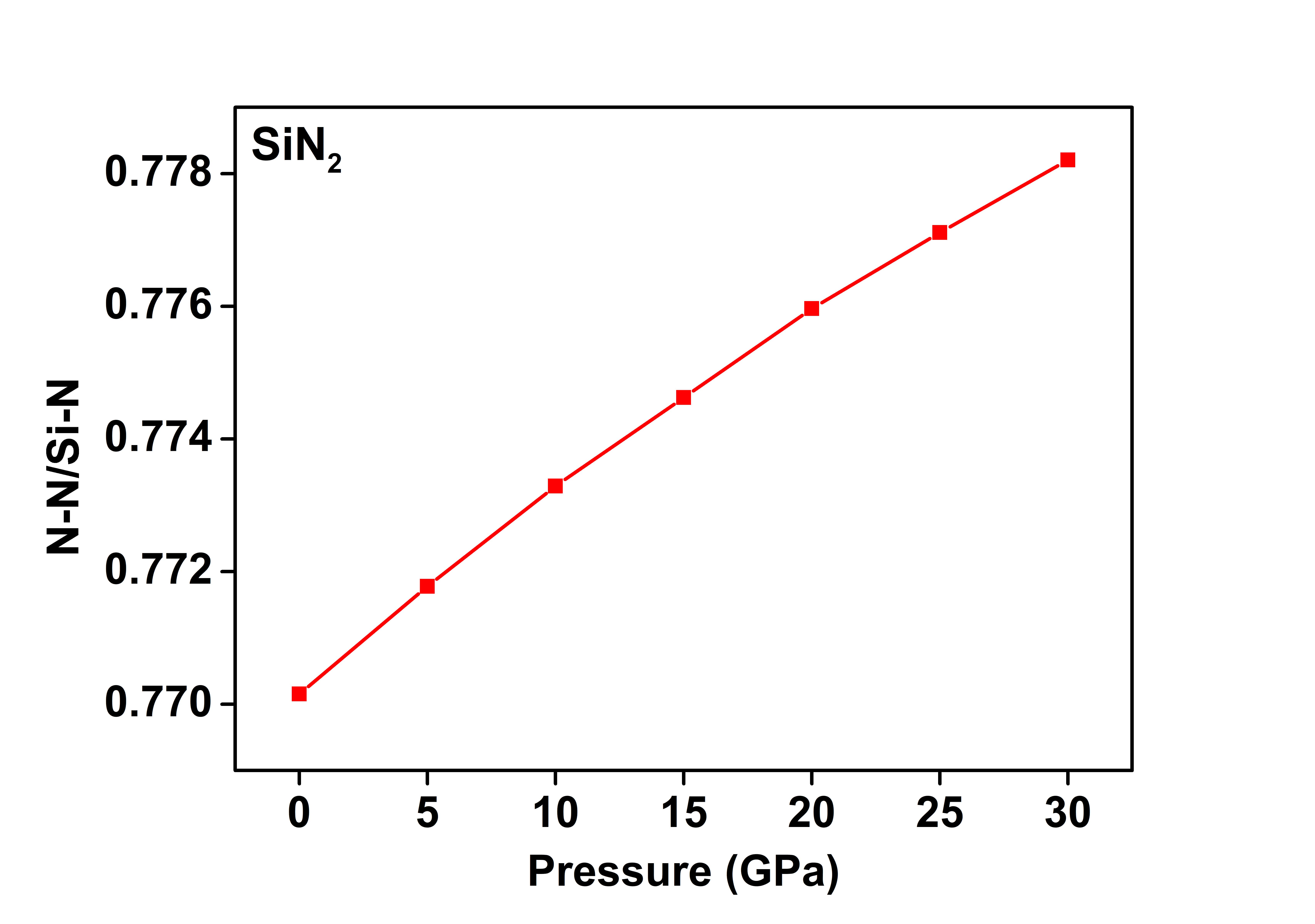}
\includegraphics[width=0.5\textwidth]{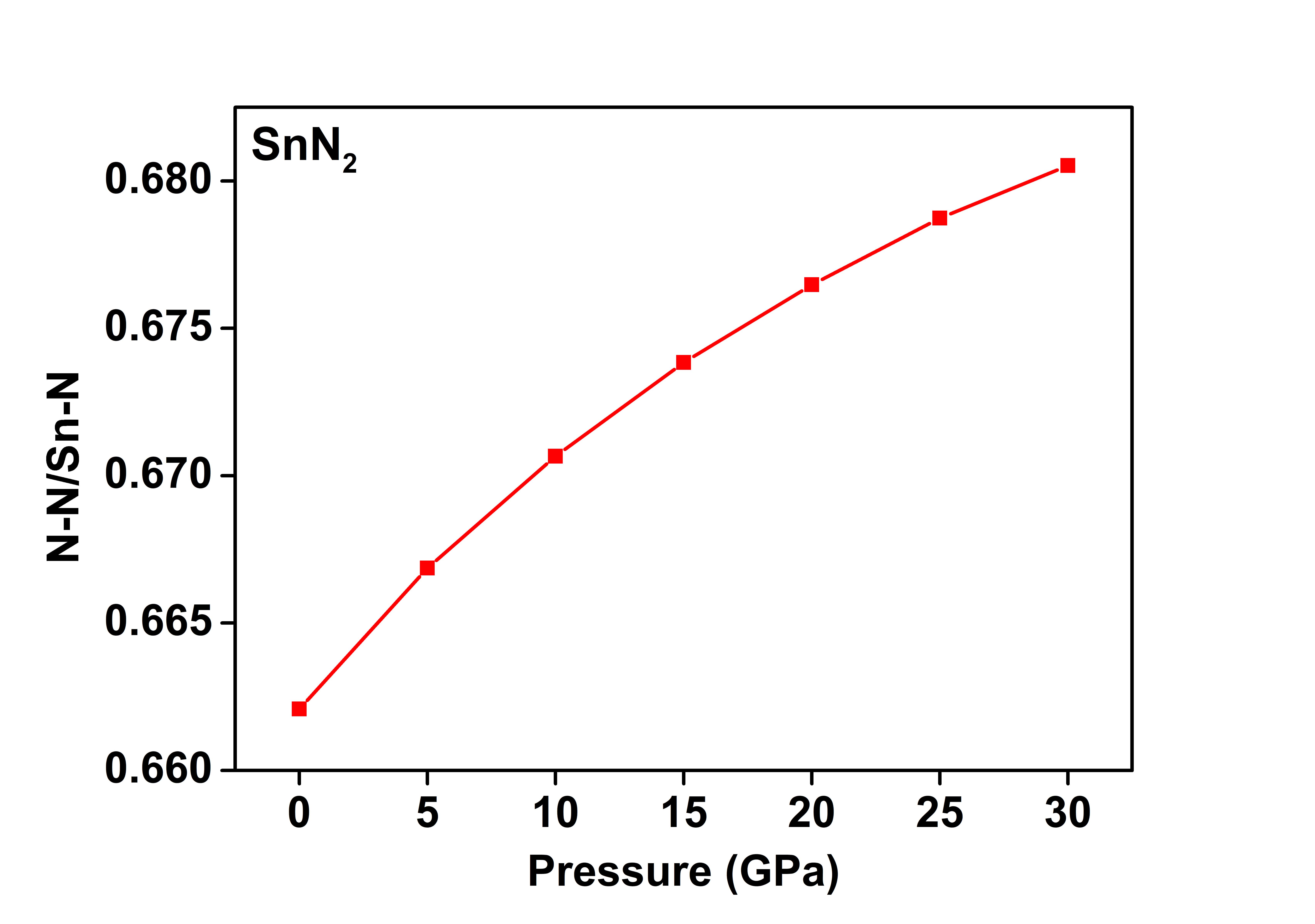}
\caption{Variation of d$_{N-N}$/ d$_{M-N}$ versus pressure for SiN$_2$ and SnN$_2$.} 
\label{fig:example_1}
\end{figure}

\begin{figure}[H]
\centering
\includegraphics[width=0.5\textwidth]{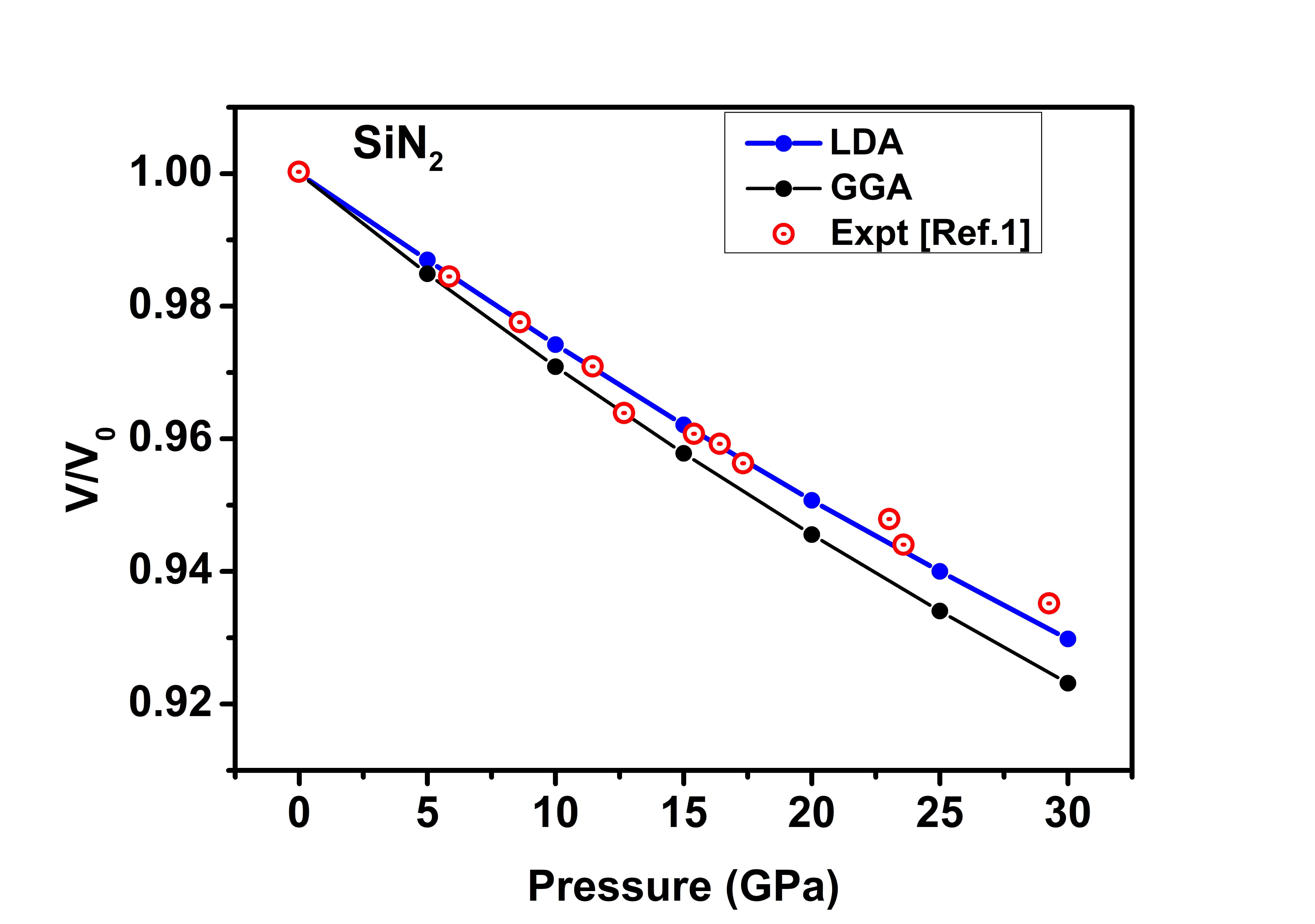}
\includegraphics[width=0.5\textwidth]{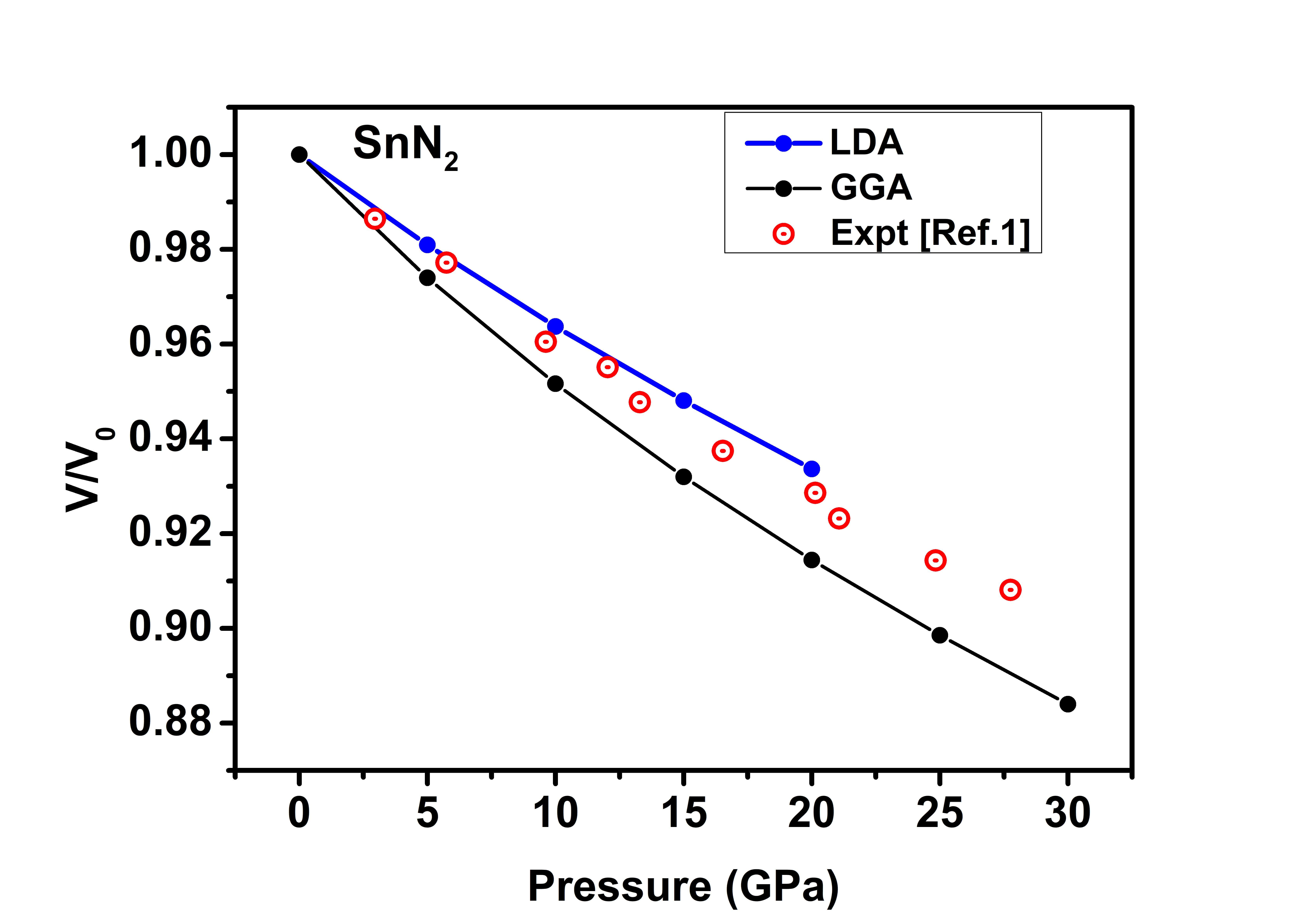}
\caption{Compressibility of SiN$_2$ and SnN$_2$ calculated using LDA and GGA functional in comparison with experimental data of Niwa et. al. \etal\cite{2017experimental}}
\label{fig:example_2}
\end{figure}
